\documentclass[11pt]{article}
\evensidemargin=\oddsidemargin

\usepackage{graphicx}
\usepackage{dcolumn}
\usepackage{bm}
\usepackage{amssymb}
\usepackage{here}

\renewcommand{\thefootnote}{\fnsymbol{footnote}}
\newcommand{\beq}{\begin{equation}}
\newcommand{\eeq}{\end{equation}}
\newcommand{\be}{\begin{equation}}
\newcommand{\ee}{\end{equation}}
\newcommand{\beqa}{\begin{eqnarray}}
\newcommand{\eeqa}{\end{eqnarray}}
\newcommand{\ba}{\begin{array}}
\newcommand{\ea}{\end{array}}

\def\O{{\cal O}}

\def\End{\end{document}}

\def\to{\rightarrow}

\def\dis{\displaystyle}
\def\f{\frac}
\def\ov{\overline}
\def\[{\left[}
\def\]{\right]}
\def\({\left(}
\def\){\right)}

\def\cut{\Lambda}

\def\aa{{\theta_1}}
\def\aab{{\theta_{12}}}
\def\ab{{\theta_2}}
\def\abc{{\theta_{23}}}
\def\ac{{\theta_3}}
\def\aac{{\theta_{13}}}
\def\D{\Delta}
\def\d{\delta}
\def\dd{\overline{\delta}}
\def\K{\kappa}
\def\KK{\overline{\kappa}}
\def\m0B{\overline{m}_0^{~}}
\def\mz{m_0^{~}}
\def\ma{{m_1^{~}}}
\def\mb{{m_2^{~}}}
\def\mc{{m_3^{~}}}
\def\deg{\circ}
\def\om{\omega}

\def\BB{\beta\beta}
\def\nuBB{0\nu\beta\beta}
\def\sq{\sqrt{2}}

\def\ts{\theta_\odot}

\def\tz{\theta_{\rm chz}}
\def\s3{s_3^{~}}
\def\ep{\epsilon}

\def\End{\end{document}}

\begin{document}                                                              
\thispagestyle{empty}
\setcounter{footnote}{1}

\begin{flushright}
{\large hep-ph/0203237}
\end{flushright}
\vspace*{15mm}

\begin{center}
{\bf {\Large \hspace*{-7mm}
Minimal~Schemes~for~Large~Neutrino~Mixings~with~Inverted~Hierarchy 
}}

\vspace*{15mm}
	{\large\sc Hong-Jian He}\,$^{\rm a}$\,\footnote{
Electronic address: hjhe@physics.utexas.edu},~~~
        {\large\sc Duane A. Dicus}\,$^{\rm a}$\,\footnote{
Electronic address: phbd057@utxvms.cc.utexas.edu},~~~
        {\large\sc John N. Ng}\,$^{\rm b}$\,\footnote{
Electronic address: misery@triumf.ca}

\vspace*{4mm}
$^{\rm a}$\,{\rm 
                 Center for Particle Physics, 
      University of Texas at Austin, Austin, Texas 78712, USA}
\\[1.7mm]
$^{\rm b}$\,{\rm 
      TRIUMF, 4004 Wesbrook Mall, Vancouver, BC V6T 2A3, Canada }
\\
\end{center}

\vspace*{65mm}
\begin{abstract}
Existing oscillation data point to nonzero neutrino masses
with large mixings.
We analyze the generic features of the neutrino Majorana mass matrix 
with inverted hierarchy and 
construct realistic {\it minimal schemes} for the neutrino mass matrix
that can explain the large (but not maximal) 
$\nu_e^{~}-\nu_\mu^{~}$ mixing of
MSW-LAM as well as the nearly maximal 
$\nu_\mu^{~}-\nu_\tau^{~}$ mixing and the
small (or negligible) 
$\nu_e^{~}\to\nu_\tau^{~}$ transition. 
These minimal schemes are quite unique and turn out to be
extremely predictive. Implications for neutrinoless double beta decay,
tritium beta decay and cosmology are analyzed.   
\\[4mm]
\hfill  [March, 2002] 
\end{abstract}

\newpage
\setcounter{page}{1}
\setcounter{footnote}{0}
\renewcommand{\thefootnote}{\arabic{footnote}}

\noindent
{\bf {\large 1. Introduction}} 
\vspace*{3mm}

The large, rather than small, neutrino mixings confirmed
by atmospheric and solar oscillation experiments 
\cite{superk,solar-new}
over the recent years have brought neutrino physics to an exciting
new era. It indicates that lepton flavor mixing is
very different from quark flavor mixing, and neutrino
mass generation may have a distinct origin from
the traditional Dirac-type Yukawa interactions for the 
charged quarks and leptons in the standard model (SM).
In fact, the neutrino masses can be naturally of Majorana
nature, generated from either a seesaw mechanism \cite{seesaw}
at high scales or a radiative mechanism around the weak scale 
\cite{zee,zad}.

The current global fit strongly favors
Mikheyev-Smirnov-Wolfenstein Large Angle Mixing (MSW-LAM) 
in which the solar mixing angle
$\theta_\odot$ 
(giving the $\nu_e^{~} \leftrightarrow \nu_\mu^{~}$ transition)
is large but significantly deviates from the
maximal value $45^\circ$,
i.e., $25^\circ \leq \theta_\odot \leq 39^\circ$ 
at 95\%\,C.L.\footnote{The maximal value $45^\deg$ is also 
excluded by the 99\%\,C.L. limit of the MSW-LAM solution, 
$24^\deg \leq \ts \leq 43^\deg$ \cite{fit,fitx,Rev}.}, 
with a central value at 
$\theta_\odot \simeq 32^\circ$ \cite{fit,fitx,Rev}.
On the other hand, the atmospheric data 
indicate a maximal mixing angle $\theta_{\rm atm}$
(representing the $\nu_\mu^{~} \to \nu_\tau^{~}$ transition), 
with the 95\%\,C.L. limit
\,$33^\circ \leq \theta_{\rm atm} \leq 57^\circ$\, and the
central value \,$\theta_{\rm atm} \simeq 45^\circ$ \cite{superk}.  
This is also supported by the K2K long baseline 
experiment\,\cite{K2K}.
The {\sc Chooz}\,\cite{CHOOZ} and Palo Verde\,\cite{PV}  
long baseline reactor experiments 
(in combination with the mass range of 
atmospheric data\,\cite{superk}) bound 
\,$\sin^2\theta_{\rm chz} \lesssim 0.04$\, at 95\%\,C.L., where the
angle $\theta_{\rm chz}$ measures the 
$\nu_e^{~} \to \nu_\tau^{~}$ transition.
Furthermore, the solar oscillations constrain the 
mass-square difference 
$\D_{\odot} =|m_1^2 - m_2^2|$ to be,
$
 1.8 \times 10^{-5}\,{\rm eV}^2 
 \leq \D_{\odot} \leq
 4.1 \times 10^{-4}\,{\rm eV}^2
$,
for MSW-LAM at 99\%\,C.L.,
while the atmospheric oscillations confine
the mass-square difference 
$\D_{\rm atm} =|m_{1,2}^2 - m_3^2|$ 
as,
$
 1.3 \times 10^{-3}\,{\rm eV}^2 
 \leq \D_{\rm atm} \leq
 5.5 \times 10^{-3}\,{\rm eV}^2
$, 
at 99\%\,C.L.
This suggests two generic patterns for the light
neutrino mass-eigenvalues, $(\ma,\,\mb,\,\mc )\geq 0$,
namely, the ``Normal Hierarchy'' (called Type-A) 
and ``Inverted Hierarchy''
(called Type-B), 
\beq
\label{eq:Type-AB}
{\rm A\!:}~~ \ma < \mb \ll \mc \,;~~~~~
{\rm B\!:}~~ \ma \sim \mb \gg \mc \,.
\eeq
Hereafter, we will focus on the Type-B scenario with
inverted mass hierarchy. Our present goal is to construct a 
realistic scheme for the neutrino Majorana mass matrix,
containing only a {\it minimal} set of parameters to describe
the neutrino data, 
especially the non-maximal solar neutrino mixing
\`{a} la MSW-LAM (which is hard\,\cite{fitx,zee-solar} to realize
in models with an approximate 
~$L_e-L_\mu-L_\tau$~ symmetry\,\cite{Barb}). 
We show that such a Minimal Scheme   
can be quite uniquely derived and is highly predictive.
Implications for neutrinoless double $\beta$ decay, 
tritium $\beta$ decay, and cosmology are analyzed.

\vspace*{6mm}
\noindent
{\bf {\large 2. 
Minimal Schemes for Neutrino Majorana Masses with Inverted Hierarchy
}}  
\vspace*{3mm}

Consider the generic $3\times3$ symmetric Majorana mass matrix
$M_\nu$
for 3 light flavor-neutrinos $(\nu_e,\,\nu_\mu,\,\nu_\tau)$,
at the weak scale and with leptons in the mass-eigenbasis,
\beq
\label{eq:Mnu}
M_\nu = 
\left\lgroup 
\ba{ccc}
m_{ee}     &  m_{e\mu}    &  m_{e\tau}   \\[2mm]
m_{e\mu}   &  m_{\mu\mu}  &  m_{\mu\tau} \\[2mm]
m_{e\tau}  &  m_{\mu\tau} &  m_{\tau\tau}
\ea
\right\rgroup .
\eeq
With extra new heavy fields integrated out, 
Eq.\,(\ref{eq:Mnu}) is the most
general description of the Majorana masses 
of three active neutrinos based upon Weinberg's
unique dimension-5 effective operator \cite{weinberg},\,
$\dis\f{{\cal C}_{ij}}{\cut}
     {L_i^\alpha} {L_j^\beta} H^{\alpha'}H^{\beta'}
 \ep^{\alpha\alpha'}\ep^{\beta\beta'}
$,\, which gives a mass term,
$\f{1}{2}\nu^T M_\nu \,\nu$,
with $M_\nu^{ij} = {\cal C}_{ij}{v^2}/{\cut}$,
where $\langle H \rangle = {v}/{\sqrt{2}}$ is the vacuum expectation
value of the SM Higgs doublet.
The mass matrix (\ref{eq:Mnu}) contains 
nine independent real parameters, which can be
equivalently chosen as three mass eigenvalues 
$(\ma,\,\mb,\,\mc )\geq 0$,\, 
three mixing angles
$(\aab ,\,\abc ,\,\aac )$,\,
and three CP-violation phases 
$(\phi,\,\phi',\,\phi'' )$ with $\phi$ the usual Dirac phase
and $(\phi',\,\phi'')$ the Majorana phases (which do not affect
the neutrino oscillation).
The neutrino mixing matrix $V\equiv UU'$ for diagonalizing $M_\nu$,
via $V^TM_\nu V = M_\nu^{\rm diag}$,
contains six parameters (three rotation angles and three phases) and can
be decomposed into a matrix $U$ (\`{a} la Cabibbo-Kobayashi-Maskawa) and 
a diagonal matrix $U'$ with only two Majorana phases, 
\beq
\label{eq:U}
U=
\left\lgroup
\ba{ccc}
c_1c_3    &   -s_1c_3         &   -s_3e^{-i\phi} \\[2mm]
s_1c_2 - c_1 s_2s_3 e^{i\phi} &  c_1c_2+s_1s_2s_3 e^{i\phi} 
& -s_2c_3  \\[2mm]
s_1s_2 + c_1c_2s_3 e^{i\phi}  &  c_1s_2 - s_1c_2s_3e^{i\phi} 
& c_2c_3
\ea
\right\rgroup
\eeq
and $U' = {\rm diag}(1,\,e^{i\phi'},\,e^{i\phi''})$\,.
Here we use the notations
$
(\aa,\,\ab,\,\ac) \equiv
(\aab ,\,\abc ,\,\aac )
$,
for convenience.
From the mass diagonalization, we can reconstruct the neutrino
mass matrix $M_\nu$ via the relation,
%
\beq
\label{eq:MnuR}
M_\nu = V^\ast M_\nu^{\rm diag}\, V^\dagger \,,
\eeq
which gives,
\beq
\label{eq:Mnuij}
\ba{lcl}
m_{ee} &\!\!=\!\!&
c_3^2
\[
c_1^2m_1+s_1^2m_2'
\] + p^2s_3^2m_3' \,,
\\[1.6mm]
m_{\mu\mu}  &\!\!=\!\!&
(s_1c_2 - \bar{p}c_1s_2s_3)^2m_1 +
(c_1c_2 + \bar{p}s_1s_2s_3)^2m_2' +
s_2^2c_3^2m_3' \,,
\\[1.6mm]
m_{\tau\tau}  &\!\!=\!\!&
(s_1s_2 + \bar{p}c_1c_2s_3)^2m_1  + 
(c_1s_2 - \bar{p}s_1c_2s_3)^2m_2' + 
 c_2^2c_3^2 m_3' \,,
\\[1.6mm]
m_{e\mu}  &\!\!=\!\!&
c_3 \[
s_1c_1c_2(m_1 - m_2') - 
\bar{p}s_2s_3(c_1^2m_1 + s_1^2m_2') + ps_2s_3m_3'
\] \,,
\\[1.6mm]
m_{e\tau}   &\!\!=\!\!&
c_3 \[
s_1c_1s_2(m_1 - m_2') + \bar{p}c_2s_3(c_1^2m_1 + s_1^2m_2')
- pc_2s_3m_3'
\],
\\[2mm]
m_{\mu\tau}   &\!\!=\!\!&
(s_1s_2 \!+\! \bar{p}c_1c_2s_3)(s_1c_2 \!-\! \bar{p}c_1s_2s_3)m_1 + 
(c_1s_2 \!-\! \bar{p}s_1c_2s_3)(c_1c_2 \!+\! \bar{p}s_1s_2s_3)m_2' 
          - s_2c_2c_3^2m_3' \,,
\ea
\eeq
where 
~$(m_2',\,m_3') \!\equiv\! 
  (m_2^{~}e^{-i2\phi'}\!\!,\, m_3^{~}e^{-i2\phi''})$~ 
and ~$p\!=\!{\bar p}^\ast \!\equiv\! e^{i\phi}$.

The precise form of the neutrino mass matrix $M_\nu$ in Eq.\,(\ref{eq:Mnu})
should be predicted by an appropriate full theory where the
mass-mechanism is known. On the other hand, 
Eq.\,(\ref{eq:MnuR}) shows how $M_\nu$ can be fully reconstructed
in terms of nine directly measurable quantities, the mass-eigenvalues,
the mixing angles and the CP-phases. 
Before knowing the underlying full theory, this suggests an important
and reliable bottom-up approach, namely, we ask: 
given the existing neutrino experiments, can we construct a simple,
realistic $M_\nu$ with only a {\it minimal set} of input parameters  
which describes all the oscillation data? 
To be concrete, we will focus on the Type-B scenario 
with inverted mass hierarchy in Eq.\,(\ref{eq:Type-AB}).\footnote{For
a very recent analysis of the normal hierarchy (Type-A) via a
bottom-up approach, see Ref.\,\cite{Yu}.}~
We will show that such a minimal scheme can be quite uniquely derived
and is highly predictive. We can, of course, further extend or elaborate
the {\it Minimal Scheme} with more fine structure and more
input parameters if that is needed to match with an underlying theory
(once specified). However, the essential structure of the Minimal Scheme and
its capability for describing the existing oscillation data\,\footnote{The
result from Liquid Scintillation Neutrino Detector
(LSND)\,\cite{LSND} awaits confirmation by the Fermilab
miniBooNE experiment\,\cite{miniB} and will not be considered
in the present study.} 
will remain in any realistic extension.

\vspace*{5mm}
\noindent
{\bf 
2.1. Minimal Scheme of Type-B1
} 
\vspace*{3mm}

The neutrino mass matrices of inverted hierarchy (Type-B) can be
classified into Type-B1 and -B2 \cite{Alt} which we will analyze in turn.
We start from the simplest,
naive mass matrix $M_{\nu 0}$ of Type-B1 \cite{Alt},
\beq
\label{eq:Mnu0}
M_{\nu 0}[{\rm B1}] =
\dis\f{\m0B}{\sqrt{2}~}
\left\lgroup
\ba{ccc}
\,0\, & \,1\, & \,1\, \\[.5mm]
\,1\, & \,0\, & \,0\, \\[.5mm]
\,1\, & \,0\, & \,0\, 
\ea
\right\rgroup  ,
\eeq
which generates a mass spectrum 
$(\ma,\,\mb,\,\mc )=
(m_1^{~},-\,m_2',\,m_3')=\m0B (1,\,1,\,0)$ and 
exact bi-maximal mixing,  $\aa=\ab=45^\circ$.   
This simple structure  (\ref{eq:Mnu0}) is motivated 
by  $L_e - L_\mu - L_\tau$ symmetry\footnote{For
some recent non-minimal approaches with certain 
$U(1)$ flavor symmetry and additional fields, see Ref.\,\cite{U1}.}.
(It was also shown to be generic
for the minimal radiative Zee-model and its various extensions
\cite{zee,zad,zeenew}.)
However, (\ref{eq:Mnu0}) is not realistic and is excluded
by the solar oscillation data since it predicts  
$\D_{\odot}=|m_1^2-m_2^2|=0$, and, 
more seriously, a maximal solar angle
$\theta_\odot = \theta_1 =45^\circ$ which is difficult to
reconcile with the MSW-LAM\,\cite{zee-solar}\cite{fitx}.
We observe that such a failure is due to the small 
but nonzero ratio of the two measured mass-square differences
$
\D_{\odot}/\D_{\rm atm} = \O(10^{-1}-10^{-2})
$ 
and a moderate angular deviation   
$(45^\circ -\theta_1)/\theta_1 
\sim (45^\circ -32^\circ)/32^\circ
\sim 0.4$, for MSW-LAM.
Therefore, it is justified to take $M_{\nu 0}$
as our zeroth order mass matrix and build in the 
necessary {\it Minimal Perturbations} 
to make a realistic Type-B1 neutrino mass-matrix
$M_\nu = M_{\nu 0} + \D M_{\nu}$. Such minimal perturbations
represent proper  $L_e - L_\mu - L_\tau$ violation effects.
What is the {\it minimal set} of extra parameters 
which we need for a realistic perturbation
$\D M_{\nu}$?  
First, we need a sizable parameter 
$\K=\O(0.5)$ to accommodate the solar angular deviation 
of $(45^\deg - \aa)/\aa \sim 0.4$; 
second, we need a small parameter 
$\d' \sim |\ma-\mb |/\m0B \lesssim \O(0.1)$ 
to account for the minor mass ratio
$\D_\odot / \D_{\rm atm} = \O(10^{-1}-10^{-2}) $;
finally, to ensure the Type-B mass spectrum (\ref{eq:Type-AB})
we should impose a condition 
$|\ma-\mb |/\m0B \sim \mc /\m0B = \O(\d')$, 
which can be naturally realized only if
we introduce an ``interplay'' parameter $\d$ lying between
$\K$ and $\d'$. In summary, to construct a realistic 
perturbation to $M_{\nu 0}$, we have to start with
three dimensionless parameters
$(\K,\,\d,\,\d')$ satisfying the proper hierarchies,
\beq
\label{eq:Kdd}
\ba{l}
1 > |\K| > |\d| > |\d'|\,,~~~~ |\K| \gg |\d'|\,,
\\[1.6mm]
\mc /\m0B \sim |\ma - \mb |/\m0B  = \O(\d') \,.
\ea
\eeq
With these, we can almost uniquely determine
the pattern of the perturbation $\D M_\nu$, and derive
the following Minimal Scheme-B1:
\beq
\label{eq:MS-B1}
M_{\nu }[{\rm B1}] =
\dis\f{\m0B}{\sqrt{2}~}
\left\lgroup
\ba{ccc}
\K &   1 &   1 \\[.8mm]
 1 & -\K & -\d \\[.8mm]
 1 & -\d & -\d' 
\ea
\right\rgroup .
\eeq
The relative sign between 11- and 22-entry is uniquely fixed
by the requirement $|\ma - \mb |/\m0B = \O(\d')$. 
Note that to affect $\theta_1$, $\K$ cannot be put in 
12- and 21-entry as $M_\nu$ is symmetric.
Another reason to arrange the 11-entry to be of $\O(\K)$
rather than $\O(\d,\,\d')$ comes from the generic
observation about the nature of $m_{ee}$  by using
the Type-B mass spectrum (\ref{eq:Type-AB}) and 
the general Eq.\,(\ref{eq:Mnuij}),
\beq
\label{eq:meeE}
\ma c_3^2 ~\gtrsim ~|m_{ee}|~ \gtrsim~ 
\ma c_3^2|\cos 2\theta_1 | \simeq \ma |\cos 2\theta_1 |, 
\eeq
where the upper [lower] bound corresponds to 
the CP-conserving values of the Majorana phase 
\,$\phi' = 0,~{\rm or,}~\pi ~
  [\pi/2,~{\rm or,}~3\pi/2]$\,. 
For solar mixing within the 95\%\,C.L. range,
$25^\deg \leq \theta_1 \leq 39^\deg $  
(MSW-LAM), we deduce the lower limit,
\beq
\label{eq:meeN}
|m_{ee}|/m_1^{~}  \approx
|m_{ee}|/\m0B ~\gtrsim~ 
(0.21-0.64) \,,
\eeq
so that we can identify 
$\sqrt{2}\,|m_{ee}^{~}|/\m0B =\O(\K)$.
It is important to note that, for {\it general} Type-B scenarios, 
the significant deviation of 
$0.15\leq (45^\deg -\theta_1)/\theta_1 \leq 0.8$
for  $25^\deg \leq \theta_1 \leq 39^\deg $ 
\cite{fit} \`{a} la MSW-LAM 
already requires a sizable $m_{ee}^{~}$ which is potentially
observable via $0\nu\beta\beta$-decay experiments 
\cite{BB-Rev}, depending on the overall scale $\m0B$\,.
As will be shown in Sec.\,3.1, due to the condition
(\ref{eq:Kdd}) and the smallness of $s_3^2$\,\cite{CHOOZ},
we can reduce the dimensionless inputs $(\K,\,\d,\,\d')$
down to a {\it single} parameter $\K$ (or, equivalently, the
solar angle $\ts\simeq\aa$). This makes our minimal
scheme-B1 highly predictive.

After a scan for all possible variations of the minimal Scheme-B1 under
the condition (\ref{eq:Kdd}), we find a few other acceptable 
minimal schemes with $M_\nu\equiv(\m0B/\!\sq)\, {\mathfrak M}$ and
${\mathfrak M}$ given by
\beq
\label{eq:MS-B1'}
\dis
\left\lgroup
\ba{ccc}
\K     &   1   &   1 \\[.8mm]
 1     & -\d'  & -\d \\[.8mm]
 1     & -\d   &  -\K 
\ea
\right\rgroup ,~~~
\left\lgroup
\ba{ccc}
\K     & 1-\d' &   1 \\[.8mm]
 1-\d' & -\K   & -\d \\[.8mm]
 1     & -\d   &   0 
\ea
\right\rgroup ,~~~
\left\lgroup
\ba{ccc}
\K     & 1-\d' &   1 \\[.8mm]
 1-\d' & 0     & -\d \\[.8mm]
 1     & -\d   &  -\K 
\ea
\right\rgroup .
\eeq
Here the first matrix has the same mass eigenvalues as 
Eq.\,(\ref{eq:MS-B1});  its rotation angles
$(\ac,\,\ab)$ contain a sign flip for the small $\O(s_3)$ terms.
The third matrix in Eq.\,(\ref{eq:MS-B1'})
is a variation of the first matrix by relocating $\d'$; 
similarly, the second matrix above is constructed
by relocating $\d'$ from  Eq.\,(\ref{eq:MS-B1}).
Hence, Eq.\,(\ref{eq:MS-B1'}) differs from 
Eq.\,(\ref{eq:MS-B1}) only by small terms of $\O(s_3,\,\d')$,
with no conceptual difference.
We will focus on the minimal scheme (\ref{eq:MS-B1}) hereafter. 
We also note that all the realistic minimal schemes we
find for the Type-B
can have at most {\it one} independent texture zero 
[cf. the above Eq.\,(\ref{eq:MS-B1'}) 
and the following Eq.\,(\ref{eq:MS-B2})
with $\xi'=0$ or $\xi=0$]. 
A recent interesting analysis\,\cite{danny} 
classified viable schemes with two independent texture-zeros,
which, as expected, do not contain Type-B schemes.

\vspace*{5mm}
\noindent
{\bf 
2.2. Minimal Scheme of Type-B2
} 
\vspace*{3mm}

The naive form of Type-B2 is defined as \cite{Alt},
\beq
\label{eq:MS-B20}
M_{\nu 0}[{\rm B2}] =
\dis {\mz}
\left\lgroup
\ba{ccc}
 1 &   0         &   0        \\[.5mm]
 0 &   {1}/{2}   &  {1}/{2}   \\[.5mm]
 0 &   {1}/{2}   &  {1}/{2}  
\ea
\right\rgroup ,
\eeq
which has a Type-B mass-spectrum 
$(m_1^{~},\,m_2^{~},\,m_3^{~}) = 
 (m_1^{~},\,m_2',\,m_3') = m_0^{~}(1,1,0)$, 
a maximal mixing angle $\ab=45^\deg$,
and vanishing $(\aa,\, \ac)$.
To be realistic,
the zeroth order matrix (\ref{eq:MS-B20}) has to be properly 
perturbed for generating the observed small but nonzero
$\D_\odot =|m_1^2-m_2^2|$ and the large (rather than
maximal) mixing  $\aa$ for MSW-LAM.
Using the fact of $m_1^{~} \simeq m_2^{~} \gg m_3^{~}$ and the general
relation (\ref{eq:Mnuij}), 
we find all the perturbations in 
$\Delta M_{\nu} = M_\nu - M_{\nu 0}$  to be of
$\O(\ma - \mb,\, \mc ,\, s_3^2m_0^{~})$ or smaller. This suggests
a {\it completely different} perturbation structure from Type-B1,
namely, we use only small perturbation parameters of 
$\O(|\ma -\mb |/m_0^{~},\, \mc /m_0^{~})$ and the large
solar mixing angle  $25^\deg \leq \aa \leq 39^\deg$
(95\%\,C.L.) can be naturally generated 
from an $\O(1)$ {\it ratio} of two
small perturbation parameters.
Inspecting the structure of $\Delta M_{\nu} = M_\nu - M_{\nu 0}$ 
for Type-B2 and using  Eq.\,(\ref{eq:Mnuij}), 
we are quite uniquely led to the following
Scheme-B2, 
\beq
\label{eq:MS-B2}
M_{\nu}[{\rm B2}] =
\dis {\mz}
\left\lgroup
\ba{ccc}
 1\!+\!\d   &   \xi      &  \xi'        \\[1mm]
 \xi        &   {1}/{2}  &  {1}/{2}     \\[1mm]
 \xi'       &   {1}/{2}  &  {1}/{2}  
\ea
\right\rgroup ,
\eeq
where we impose
\beq
\dis
 \f{\,\mc\,}{\mz} ~\lesssim~ \f{~|\ma\!-\!\mb |~}{\mz}  
 ~=~ \O(\d,\xi,\xi') ~\,\ll\,~ 1 \,.
\eeq
We can define a truly {\it Minimal Scheme-B2}\,  by setting ~$\xi'=0$\,.\,
Choosing $\xi'=\xi$ will give,
$\ac=\O(\d^2,\xi^2,s_3^2,\d \s3,\xi \s3 )\simeq 0$, 
and is consistent with the
{\sc Chooz} bound\,\cite{CHOOZ}. But Type-B2
generally has negligible $\s3$ even for nonzero $\xi'\neq \xi$.
In Eq.\,(\ref{eq:MS-B2}), 
there is no need to perturb the $2\times2$ block of 
$\nu_\mu-\nu_\tau$ as the maximal mixing is favored by
the atmospheric data\,\cite{superk}; also the $2\times2$ block of 
$\nu_e-\nu_\mu$ invokes two small perturbations so that
an $\O(1)$ ratio is generated to explain the non-maximal
solar mixing (MSW-LAM). 
Unlike the Type-B1, the scheme-B2 has 
larger ~$m_{ee}^{~} = m_0^{~}(1+\d)\simeq m_0^{~}$~ at the {\it zeroth} 
order and is more sensitive to the $\nuBB$ experiments. 

\vspace*{7mm}
\noindent
{\bf {\large 
3. Analysis of the Minimal Schemes and Predictions 
   for Neutrino Oscillation 
}} 
\vspace*{3mm}

In this section, we systematically solve the diagonalization
equations in (\ref{eq:Mnuij}) for the Minimal Scheme-B1
(\ref{eq:MS-B1}) and -B2 (\ref{eq:MS-B2}) [$\xi'=0$] 
with CP-conservation. We then study their predictions for the
neutrino oscillations.

\vspace*{3mm}
\noindent
{\bf 3.1. Analyzing the Minimal Scheme of Type-B1 
}
\vspace*{2mm}

The parameters $(\K,\,\d)$ will be retained up to all orders
without approximation. 
But, from the solar and {\sc Chooz} oscillation data, 
it is justified to treat the small parameters $(\d',\,\s3 )$ as 
perturbations to first power and ignore terms
of $\O({\d'}^2,\,s_3^2) \lesssim \O(10^{-2})$ or smaller.
As will be shown below, the expansion of $\s3$
also plays a key role for eliminating $\d'$ from inputs.

From Eq.\,(\ref{eq:MS-B1}),
we deduce the mass-eigenvalues of $M_\nu$, 
up to $\O(\ov{\d'})$,
\beq
\label{eq:m123}
\dis 
m^{~}_{1,2} = \mz \[1\mp\f{1}{2}\(1-\f{x}{\,2+\om\,}\)\ov{\d'}\] ,~~~~  
        \mc = \mz \f{x}{\,2\!+\!\om\,}\,\ov{\d'} \,,
\eeq
where we expand $\ov{\d'}$ to first order and define,
\beq
\label{eq:barKdd}
\ba{l}
\dis
\KK      \equiv \f{\K}{~\ov{\om}~} \,,~~~
\dd      \equiv \f{\d}{~\ov{\om}~} \,,~~~
\ov{\d'} \equiv \f{\d'}{~\ov{\om}~}\,,~~~
\ov{\om} \equiv \sqrt{2+\om\,} 
=\sqrt{\f{2}{1-\KK^2-\dd^2\,}}
\,,
\\[-1.2mm]
\\[-1.2mm]
\dis
\om \equiv \K^2\!+\!\d^2 
    =\f{\,2(\KK^2\!+\!\dd^2)\,}{\,1-(\KK^2\!+\!\dd^2)\,} ,~~~~
\mz \equiv \m0B\sqrt{1\!+\!\f{\om}{2}\,}  \,.
\ea                      
\eeq
The parameter $x=\O(1)$ in Eq.\,(\ref{eq:m123}) will be
determined by the consistency condition,
\beq
\label{eq:x}
\(1-\d^2\)\K-2\d+\(1+\K^2\)\d' = x\,\d'  \,,
\eeq
due to the requirement $\mc /\m0B = \O(\d')$ in 
Eq.\,(\ref{eq:Kdd}).  
With the definition of Eq.\,(\ref{eq:barKdd}), we can 
rewrite the neutrino mass matrix (\ref{eq:MS-B1}) 
scaled by $\mz$,
\beq
\label{eq:MS-B1X}
M_{\nu }[{\rm B1}] =
\dis {\mz}
\left\lgroup
\ba{ccc}
\!\!\KK           & ~~\ov{\om}^{\,-1} &  ~~\ov{\om}^{\,-1}  \\[.8mm]
~\ov{\om}^{\,-1}  & -\KK & -\dd      \\[.8mm]
~\ov{\om}^{\,-1}  & -\dd & -\ov{\d'}
\ea
\right\rgroup .
\eeq
From Eq.\,(\ref{eq:m123}), we deduce
\beq
\label{eq:B1massR}
\dis\f{\D_\odot}{\,\D_{\rm atm}\,}
\,=\, \f{|m_1^2-m_2^2|}{~|m_{1,2}^2-m_3^2|~}  
\,\simeq\, 2\[1-\f{x}{2+\om}\]\ov{\d'}  \,.
\eeq

With the mass-eigenvalues given, we can then solve for the mixing
angles by substituting $M_\nu$ 
[cf. Eq.\,(\ref{eq:MS-B1}) or Eq.\,(\ref{eq:MS-B1X})]
into the six diagonalization equations in (\ref{eq:Mnuij})
and expanding $(\d',\,\s3 )$ systematically to 
first order. At this order, we find that 
only five of the six equations are independent.
Note that we have three dimensionless parameters $(\K,\,\d,\,\d')$ in 
$M_\nu$ (in which the overall scale $\m0B$ is irrelevant
to the diagonalization) 
and three mixing angles $(\aa,\,\ab,\,\ac)$. 
Hence, from the five equations, we can
solve five out of the six parameters 
as functions of a {\it single} 
dimensionless input parameter which will be chosen as the
angle $\theta_1$ (measured in the solar oscillation). 
To explicitly understand this nontrivial reduction 
of input parameters, we first note that
even though we have three dimensionless 
inputs $(\K,\,\d,\,\d')$ in (\ref{eq:MS-B1}),
the condition $\mc /\m0B = \O(\d')$ in (\ref{eq:Kdd}) 
[or, (\ref{eq:x})] relates $\K$ and $\d$
at zeroth order of $\d'$ so that only two inputs
among $(\K,\,\d,\,\d')$ are independent under the expansion of $\d'$.
Then, we summarize two relevant relations derived from 
Eq.\,(\ref{eq:Mnuij}) [and Eq.\,(\ref{eq:MS-B1X})],
\beq
\label{eq:redu}
\ba{lcl}
{\dd}^2 &=& (1-r){\KK}\,{\ov{\d'}}  +  \O(s_3^2,\,{\ov{\d'}}^2) \,, \\[1.5mm]
{\KK}   &=& 2{\dd}-(1-2r)\ov{\d'} +  \O(s_3^2) \,,
\ea
\eeq
with  
%
$\dis r\equiv \f{x}{2+\om} \,$.
%
Now we see that the {\it absence} of $\O(\s3 )$ term in 
Eq.\,(\ref{eq:redu}) and the smallness 
of $s_3^2 \,(\lesssim 0.04$ \cite{CHOOZ}) 
lead us to have three constraints
[two in Eq.\,(\ref{eq:redu}) and one in Eq.\,(\ref{eq:x})]
among the four parameters $(\KK,\,\dd,\,\ov{\d'},\,x)$\,. 
This feature remains if we include higher order terms
via {\it iteration.}
This makes our scheme end up with a {\it single} input for 
all mixings and thus extremely predictive.
After a lengthy and careful derivation, we arrive at the
following complete set of solutions of our Minimal Scheme-B1, up to
$\O(\ov{\d'},\,\s3 )$\,,
\beq
\label{eq:solution}
\ba{l}
\dis 
\KK = \f{8}{9}\cos 2\theta_1 \,,~~~
\dd = \f{1}{2}\KK \,,~~~
\ov{\d'} = \f{\KK}{4(1-r)} \,,
\\[5mm]
\dis
\theta_2 
= \f{\pi}{4}\!-\!\f{3\!-\!4r}{8(1\!-\!r)}\KK^2 ,~~
\theta_3 \simeq \s3 
= \f{3\!-\!4r}{8(1\!-\!r)}\KK\sqrt{1\!-\!\KK^2},
\ea
\eeq
where we have,
\beq
\label{eq:solutonx}
\hspace*{-2mm}
\dis
2+\om = \f{2}{1-\f{5}{4}\KK^2} \,,~~~
\dis
r = \f{1\!-\!\f{5}{4}\KK^2}{\,2(1-\KK^2)\,} \,,~~~
x = 
        \f{1}{\,1-\KK^2\,} \,.
\eeq
Finally, inputting the solar angle $\theta_\odot (\simeq \aa)$, 
we deduce the following numerical predictions,
\beq
\label{eq:nout}
\ba{l}
25^\deg \leq \theta_1 \leq 39^\deg
,~~[{\rm Input~of~MSW\!\!-\!\!LAM,}~ 95\%{\rm C.L.}];
\\[2mm]
39.8^\deg \leq \theta_2 \leq 44.5^\deg ,~~~
0.13 \geq \s3 \simeq \theta_3 \geq 0.046 ,
\\[2mm]
0.57 \geq \KK \geq 0.19   ,~~
0.29 \geq \dd \geq 0.092  ,~~
0.25 \geq \ov{\d'} \geq 0.091 ;
\\[2.5mm]
\dis
0.28 \geq \f{\D_\odot}{\D_{\rm atm}} \geq 0.095 \,,
\ea
\eeq
and also
%
$~
 0.44 \leq r \leq 0.50\,,~\,
 1.48 \geq x \geq 1.01\,.\, 
$
%
The results in Eq.\,(\ref{eq:nout})
agree well with the oscillation 
data\,\cite{superk,solar-new,fit,fitx,CHOOZ}, 
i.e., 
$33^\deg \leq \ab \simeq \theta_{\rm atm} \leq 57^\deg$\,
and
\,$\s3 = \sin\theta_{\rm chz} \lesssim 0.2$\, at 95\%\,C.L., and
$
4.2\,[3.3]\times 10^{-3}
\,\leq \D_\odot / \D_{\rm atm} \leq\, 
0.17\,[0.32]$\,
at 95\%[99\%]\,C.L.
The on-going KamLAND experiment\,\cite{Kam} 
will more precisely test the MSW-LAM parameter space 
(though it will not be sensitive to $s_3^{~}$ \cite{Kam}).
We have checked the numerical accuracy of the above
solutions by substituting Eq.\,(\ref{eq:nout}) 
back into Eqs.\,(\ref{eq:Mnuij}) and (\ref{eq:MS-B1}) and 
evaluating the difference of the two sides in each equation. 
We find that the difference (uncertainty)
is always less than  \,$0.015\,(0.0009)$\, for 
\,$\aa=25^\deg \,(39^\deg )$.\,
Thus, our systematical expansion works well
up to $\O(\ov{\d'},\,\s3 )$, as expected, 
since the ignored terms  
are of ~$\O(\ov{\d'}^2,\,s_3^2)$~ and become smaller for larger $\aa$
as shown in  Eq.\,(\ref{eq:nout}).
For Type-B schemes, the mass scale
$\mz$ is generally bounded by
\beq
\label{eq:m0-bound}
\hspace*{-2mm}
0.036\,{\rm eV} \,\leq\, m_0^{~} \!\simeq\! m_{1,2}^{~} 
                      \!\simeq\! \D_{\rm atm}^{1/2} 
                \,\leq\, 0.074\,{\rm eV} ,
~~~~[{\rm  99\%\,C.L.}].
\eeq
%

\noindent
{\bf 3.2. Analyzing the Minimal Scheme of Type-B2
}
\vspace*{3mm}

We now turn to the minimal scheme-B2 in (\ref{eq:MS-B2})
with $\xi'=0$,
which has the mass-eigenvalues, up to $\O(\d,\,\xi)$, 
\beq
\label{eq:B2-mass}
%
\dis
m_{1,2}^{~} = \mz \[1\!+\!\f{\d}{2}\!\pm \!\f{1}{2}\sqrt{\d^2\!+\!2\xi^2}\],~~~~
\mc = 0~.
\eeq
Substituting (\ref{eq:MS-B2}) into (\ref{eq:Mnuij}), 
expanding up to $\O(\d,\,\xi,\,s_3^{~})$ and using (\ref{eq:B2-mass}), 
we derive the solutions,
\beq
\label{eq:sol-B2}
\ba{l}
\dis
\theta_1 = \f{\pi}{4} \!-\!
  \f{1}{2}\arcsin\f{\d}{\sqrt{\d^2+2\xi^2\,}\,} \,,~~~
\theta_2 = \f{\pi}{4} \,,~~~  
\theta_3 =\! -\f{\xi}{\sq\,}\,,~~~
(\d,\,\xi)>0\,;
\\[6mm]
{\rm and}~~~
\dis
\f{\D_\odot}{\,\D_{\rm atm}} =
\f{|m_1^2-m_2^2|}{~|m_{1,2}^2-m_3^2|~}  
= 2\sqrt{\d^2\!+\!2\xi^2\,} \,.
\ea
\eeq
We see that $\ab$ is maximal at this order.
The sizable deviation of
~$\theta_1 - \dis\f{\pi}{4}$~ is indeed naturally generated by 
an $\O(1)$ {\it ratio} of two small parameters $(\d,\,\xi)\ll 1$.
[Allowing \,$\xi' \neq 0$,\, 
the corresponding formulas for Eqs.\,(\ref{eq:B2-mass})-(\ref{eq:sol-B2})
can be directly obtained by the simple replacements,
$\,\xi\,\to\,\xi-\xi'\,$ for $\ac$ and
$\,\xi\,\to\,\xi+\xi'\,$ for all other quantities.]
Using the inputs for LAM \cite{fit,fitx,Rev},
~$25^\deg \leq \theta_\odot \simeq \aa \leq 39^\deg $~   and 
~$4.2\times 10^{-3}\leq \D_\odot/\D_{\rm atm} \leq 0.17$~ 
at 95\%\,C.L., we deduce,
\beq
1.2\, \geq \,\d/\xi \,  \geq\, 0.3\,, ~~~~
1.1\times 10^{-3}\, \leq  \,\xi\,  \leq \,0.06\,.
\eeq
Thus we have
\beq
8\times 10^{-4} \leq\, -\ac~ \leq \,0.04 \,.
\eeq
The mass scale $\mz\simeq \D_{\rm atm}^{1/2}$ is bounded 
as in Eq.\,(\ref{eq:m0-bound}).
As mentioned above, allowing nonzero \,$\xi' = \O(\xi)$,\, 
we can derive,
\beq
s_3^{~} \,\simeq\, \ac \,=\, \dis \f{~\xi'-\xi~}{\sq}\,,
\eeq
which remains of the same order. 
Hence, $|\ac|\lesssim \O(10^{-2})$ generally holds
for Type-B2, implying  negligible CP-violation  
from the Dirac phase $\phi$\,.
It has been shown\,\cite{Barger} 
that combining the data from two
near-future long baseline accelerator experiments\,\cite{MI-IC}, 
the Main Injector Neutrino Oscillation Search (MINOS)
and the Imaging Cosmic And Rare Underground Signals (ICARUS),
may place a 95\%\,C.L. lower bound, $\ac \geq \O(0.05)$
(when $\theta_3$ lies within their combined sensitivity),
which could possibly discriminate Type-B2 from Type-B1.

\vspace*{7mm}
\noindent
{\bf {\large 4. Implications for Neutrinoless Double
$\beta$-Decay, Tritium $\beta$-Decay }}\\
{\bf {\large \hspace*{5mm}
and Cosmology
}}
\vspace*{2mm}

The oscillation data may already give a strong hint on the neutrino
mass scale [cf. Eq.\,(\ref{eq:m0-bound})]
so long as the neutrino masses exhibit the hierarchy structure
[cf. Eq.\,(\ref{eq:Type-AB})], but the possibility of three nearly
degenerate neutrinos 
($\ma \sim \mb \sim \mc$) could allow a higher scale.  
Hence, the laboratory experiment on neutrinoless double 
$\beta$-decay ($\nuBB$)\,\cite{BB-Rev,BB-new0,BBrefnew}
is indispensable to pin down the absolute
mass scale, as well as the Majorana nature of active neutrinos.
For the Minimal Scheme-B1 and -B2, we have,
\beq
\label{eq:mee}
m_{ee}^{~}[{\rm B1}] =\KK \mz \,,~~~
m_{ee}^{~}[{\rm B2}] \simeq \mz \,.
\eeq
Thus, using Eqs.\,(\ref{eq:mee}) and (\ref{eq:m0-bound}), 
we derive, at 99\%\,C.L., 
\beq
\label{eq:B1B2-MR}
0.014\,{\rm eV} 
\leq m_{ee}^{~}[{\rm B1}] \leq  
0.029\,{\rm eV} \,,~~~~
0.036\,{\rm eV} \leq m_{ee}^{~}[{\rm B2}] \leq  0.074\,{\rm eV} \,,
\eeq
where we input the central value of solar fit (LAM),  
$\theta_1\simeq 32^\deg$,  for Type-B1.
The $m_{ee}^{~}[{\rm B2}]$ is already sensitive to the current
$\nuBB$ measurement\,\cite{BB-new}. 
The experiments of $\nuBB$ decay\,\cite{BB-Rev},
such as the on-going {\sc Nemo3} and the upcoming {\sc Cuore},
can probe \,$|m_{ee}^{~}|\sim 0.1$\,eV, 
while the near-future measurements at
{\sc Genius}, {\sc Exo}, {\sc Majorana} and {\sc Moon} 
aim at a sensitivity of 
\,$|m_{ee}^{~}| \sim 0.01$\,eV, 
which is decisive for testing the whole mass range 
(\ref{eq:B1B2-MR}) of Type-B1 and -B2 schemes.

Tritium $\beta$-decay requires\,\cite{singleB},
~$m_{\nu_e}^{~} < 2.2$\,eV,\, at 95\%\,C.L., where 
~$m_{\nu_e}^{~} \equiv (M_\nu^\dag M_\nu)_{ee}^{1/2}
           \simeq m_{1,2}^{~} \simeq \mz$~ for Type-B.
This is well above the range given in Eq.\,(\ref{eq:m0-bound}).
The sensitivity of $H^3$ $\beta$-decay
could eventually reach \,$m_{\nu_e}^{~} \sim 0.5$\,eV 
\cite{KA}.

The latest cosmology measurements of the power spectrum for
the Cosmic Microwave Background (CMB), Galaxy Clustering and  
Lyman Alpha Forest \cite{cos} put a 95\%\,C.L.
upper bound on the neutrino masses, 
$\sum_j m_j^{~} \leq 4.2$\,eV.
This gives, for our Type-B schemes,  
\,$\ma \simeq \mb \simeq \mz \leq 2.1$\,eV,\, which
is about the same as the tritium $\beta$-decay bound.
The newest analysis\,\cite{2dF} from the 2dF Galaxy Redshift Survey 
arrives at an upper bound,
$\sum_j m_j^{~} \leq 2.2$\,eV, 
which results in, 
\,$\ma \simeq \mb \simeq \mz \leq 1.1$\,eV,\, for Type-B schemes.
Stronger constraints of 
\,$\sum_j m_j \lesssim 0.4$\,eV\,
are expected from the forthcoming 
Microwave Anisotropy Probe (MAP) and {\sc Planck} 
satellite experiments\,\cite{MAP1,MAPx}.

It is interesting to note 
that the neutrino mass scale may 
also be determined from the so-called $Z$-bursts \cite{ZBU0} due to
the resonant annihilation of ultra high energy neutrinos 
with cosmological relic (anti-)neutrinos into $Z$ bosons
(whose decay produces protons and photons). 
In the most plausible case 
where the ordinary cosmic rays are protons of extragalactic origin, 
the required neutrino mass range is \cite{ZBU}, 
\beq
0.01\,[0.02]\,{\rm eV} 
\,\leq\, m_\nu^{~} ({\rm heaviest}) \,\leq\, 
3.0\,[2.1]\,{\rm eV},~~~~
{\rm at  \,~99\%\,C.L. \,[95\%\,C.L]},
\eeq
which is compatible
with the neutrino oscillation bound (\ref{eq:m0-bound}) for 
the inverted mass hierarchy.

\vspace*{7mm}
\noindent
{\bf {\large 5. Conclusions  }}
\vspace*{2mm}

In this study, we have considered two essential and distinct scenarios
for the neutrino Majorana mass matrix with inverted hierarchy,
called Type-B1 [cf. Eq.\,(\ref{eq:Mnu0})] 
and -B2 [cf. Eq.\,(\ref{eq:MS-B20})].
For Type-B1, we start with the form of Eq.\,(\ref{eq:Mnu0}) at zeroth 
order and perturb it into the realistic form of Eq.\,(\ref{eq:MS-B1}) 
with three parmeters $(\K,\,\d,\,\d')$ under the hierarchy
(\ref{eq:Kdd}) that is necessary to correctly predict the 
oscillation data, especially, 
the non-maximal solar neutrino mixing of MSW-LAM.
The sizes of 
$\d'\sim |\ma - \mb |/\mz \sim \D_\odot/\D_{\rm atm} \ll 1$ 
\cite{solar-new}
and $\s3 = \sin\tz \ll 1$ \cite{CHOOZ}
justify the expansion of $(\d',\,\s3 )$, which enables us to reduce the
number of inputs down to a {\it single} parameter $\K$, or, 
equivalently, $\ts (\simeq \aa)$. 
Thus, using only the measured solar angle $\theta_\odot$ as input,
we predict the atmospheric mixing angle, 
$\theta_{\rm atm}(\simeq \ab )$, the
value of $ \theta_{\rm chz} (=\ac)$, and the mass ratio
$\D_\odot /\D_{\rm atm}$, in complete agreement with the existing data.
We also note that the Minimal Scheme-B1 in Eq.\,(\ref{eq:MS-B1}) points to a 
generic way for naturally extending the minimal Zee-model\,\cite{zee} in which
$M_\nu$ exhibits the following structure,
\beq
\label{eq:Zee}
M_\nu^{\rm Zee} = 
\left\lgroup 
\ba{ccc}
0              &  m_{e\mu}^{~}    &  m_{e\tau}^{~}   \\[2mm]
m_{e\mu}^{~}   &  0               &  m_{\mu\tau}^{~} \\[2mm]
m_{e\tau}^{~}  &  m_{\mu\tau}^{~} &  0
\ea
\right\rgroup ,
\eeq
where the pattern 
$\,m_{e\mu}^{~}\simeq m_{e\tau}^{~} \gg m_{\mu\tau}^{~}\,$
can be realized\,\cite{zeenew},
which ensures an approximate $L_e-L_\mu-L_\tau$ symmetry.
The necessity of modifying Eq.\,(\ref{eq:Zee}) in the minimal Zee-model 
for accommodating the MSW-LAM was noted recently\,\cite{zee-solar}.  
Our minimal construction of Scheme-B1 in Eq.\,(\ref{eq:MS-B1})
demonstrates a generic way to extend $M_\nu^{\rm Zee}$ 
under an appropriate perturbation [cf. Eq.\,(\ref{eq:Kdd})]. 
The choice of ~$m_{ee}^{~}=(\m0B/\sq)\K =\mz \,\KK$~ 
is due to the general
observation in Eq.\,(\ref{eq:meeE}) for Type-B1
and a sizable \,$\cos 2\aa\in (0.21-0.64)$\, for
\,$25^\deg \leq \aa \leq 39^\deg$\, (95\%\,C.L., MSW-LAM);
while \,$m_{\mu\mu}^{~}\simeq -m_{ee}^{~}$\, is enforced 
by the Type-B mass-spectrum, 
\,$m_{1,2}^{~} \gg |\ma - \mb | \sim \mc$.\,
The $\K$ terms in $M_\nu[{\rm B1}]$ represent the
generic leading modification to the minimal Zee-model (\ref{eq:Zee}).

For Type-B2, we start with the leading order mass matrix 
(\ref{eq:MS-B20}) and find the perturbation structure
in Eq.\,(\ref{eq:MS-B2}) based on the general relations
in Eq.\,(\ref{eq:Mnuij}) and 
the smalless of $\theta_{\rm chz} (=\ac)$\,.
The realistic Minimal Scheme-B2
contains only two small parameters $(\d,\,\xi)\ll 1$, 
defined as in Eq.\,(\ref{eq:MS-B2}), when \,$\xi'=0$\,.\,   
In contrast to Typy-B1, the non-maximal solar mixing
angle $\theta_\odot$ is naturally accommodated by a ratio of
two small parameters, $\d/\xi =\O(1)$, 
while the atmospheric mixing angle remains maximal.
Using the measured values of solar angle $\ts$ and mass ratio
$\D_\odot /\D_{\rm atm}$, we derive the ranges for $\ac (=\tz )$ 
and the perturbation parameters $(\d,\,\xi)$. 
The angle $\ac$ is found to be of $\O(10^{-2})$ or smaller.  
Combining the data from both MINOS and ICARUS experiments\,\cite{MI-IC} 
may reach the sensitivity\,\cite{Barger}
to discriminate between the minimal Type-B1 and Type-B2 schemes.

The overall neutrino mass scale $m_0$ for the inverted mass hierarchy
is quite uniquely fixed by the
atmospheric neutrino data on the mass-squared difference 
$\D_{\rm atm}=|m_{1,2}^2-m_3^2|\simeq m_{1,2}^2\simeq m_0^2$ 
[cf. Eq.\,(\ref{eq:m0-bound})].
Thus, the mass matrix of our minimal scheme-B1 or -B2 is 
known and highly predictive. Some implications of
the Type-B1 and -B2 minimal schemes for
the neutrinoless double $\beta$-decay, tritium $\beta$-decay and
cosmology are given above.
\\

\noindent
{\bf Note Added:}
After the submission of this work, a new announcement\,\cite{SNO2} 
from the Sudbury Neutrino Observatory (SNO) collaboration 
appeared on April 20, 2002, which further
confirms the MSW-LAM as the best solution 
to the solar neutrino oscillations.

\vspace*{4mm}
\noindent
{\bf {\large Acknowledgments}}
\\[1.7mm]
We are happy to thank Vernon Barger, Danny Marfatia 
and Alexei Yu$.$ Smirnov for valuable discussions.
This work was supported in part by U.S. Department of Energy under
grant DE-FG03-93ER40757 and
by Natural Science and Engineering Research Council of Canada.



\end{document}